# Learning When to Take Advice: A Statistical Test for Achieving A Correlated Equilibrium

**Greg Hines**
Cheriton School of Computer Science
University of Waterloo
Waterloo, Canada
ggdhines@cs.uwaterloo.ca

**Kate Larson**
Cheriton School of Computer Science
University of Waterloo
Waterloo, Canada
klarson@cs.uwaterloo.ca

## Abstract

We study a multiagent learning problem where agents can either learn via repeated interactions, or can follow the advice of a mediator who suggests possible actions to take. We present an algorithm that each agent can use so that, with high probability, they can verify whether or not the mediator's advice is useful. In particular, if the mediator's advice is useful then agents will reach a correlated equilibrium, but if the mediator's advice is not useful, then agents are not harmed by using our test, and can fall back to their original learning algorithm. We then generalize our algorithm and show that in the limit it always correctly verifies the mediator's advice.

## 1 Introduction

In settings where agents repeatedly interact with each other (for example, through a repeated game), there are great opportunities for learning since agents are able to adapt their strategies given the history of play. This problem has garnished a lot of attention from several research communities, including the AI community and the game theory community. While many criteria have been proposed for measuring the success of learning approaches, one commonly used measure is whether the agents learn how to best-respond to the strategies being played by the others. That is, does the learning process converge to an equilibrium.

In this paper we study the problem of agents interacting with each other in a repeated game setting, but we introduce a third party *mediator* or *advisor* who makes strategy suggestions to the agents. Ideally, by following the suggestions of the mediator, agents will be able to learn how to play against each other, possibly even reaching mutually beneficial outcomes which would not have been possible without the mediation. That is, our goal is for the agents to learn and adapt so that they find a correlated equilibrium [1].

However, a mediator is only useful if it can make good suggestions. Even if a mediator tries to make good suggestions it may be prevented by coding errors, memory limitations, *etc*. For an agent to accept a mediator's suggestions, there must be some way for the agent to verify that the suggestions are reasonable. A mediator might not be willing to share its code with the agents, or be aware of its own limitations. Therefore, for a truly robust system, the agents themselves must have a way of checking the mediator's suggestions.

Thus, this paper introduces a statistical test based on hypothesis testing that, with high probability, can verify the mediator's suggestions. While hypothesis testing has been proposed in the multiagent learning literature as a tool that agents might use to learn how to play Nash equilibria [5], to the best of our knowledge it has never been applied for validating a mediator's advice. Based on our test, we propose an algorithm that allows agents to converge to the mediator's suggestion if it is a correlated equilibrium and otherwise, in the limit, be no worse off for having used our algorithm. We then generalize this algorithm to a more theoretical setting where we show that with probability one, in the limit, our test will always be able to correctly verify the mediator's suggestions. This provides a method for achieving convergence to a specific correlated equilibrium.

## 2 Background

In this section we introduce the key concepts and assumptions used in this paper.

A $n$-agent *stage game* is a tuple $G = \langle N, A = A_1 \times \ldots \times A_n, u_1, \ldots, u_n \rangle$, where $N = \{1, \ldots, n\}$ is the set of agents, $A_i$ is the set of possible actions for agent $i$ and $A$ is the set of possible joint actions, and $u_i : A \to \mathbb{R}$ is the utility function for agent $i$. Without loss of generality, we assume that all utilities are greater than or equal to 0. A specific action for agent $i$ is $a_i \in A_i$, and a joint action is $a = (a_1, \ldots, a_n)$. We assume that $A$ is public knowledge but the agents' utility functions are private.

Each agent chooses its actions according to some *strategy*. A strategy for agent $i$, $\sigma_i$, is a probability distribution over $A_i$, stating with what probability the agent will play each possible action. The set of all possible strategies for agent $i$ is $\Sigma_i$. The vector $\sigma = (\sigma_1, \ldots, \sigma_n)$ is a strategy profile which specifies a strategy for each agent and $\Sigma$ is the set of all possible strategy profiles. We use $\sigma_{-i}$ to denote $(\sigma_1, \ldots, \sigma_{i-1}, \sigma_{i+1}, \ldots, \sigma_n)$.

Given a strategy profile $\sigma$, we define the *expected utility* for agent $i$ as

$$u_i(\sigma) = \sum_{a=(a_1,\ldots,a_n) \in A} u_i(a) \Pi_{j=1}^n \sigma_j(a_j). \quad (1)$$

Each agent's utility is dependent not just on its own actions, but also on the actions taken by all other agents. We assume agents are *rational*, *i.e.*, given $\sigma_{-i}$, agent $i$ will choose a strategy which maximizes its expected utility.

In our model we introduce a third-party mediator, $\mathcal{M}$. The mediator knows the utility functions for all agents, but is not affected by the game's outcome. Instead $\mathcal{M}$ makes suggestions to each agent as to what action it should take, where these suggestions are instantiations of a correlated strategy.

**Definition 1.** *A correlated strategy, $\sigma_A$, is a probability distribution over $A$. We let $s \in A$ denote an instantiation of $\sigma_A$. The conditional correlated strategy $\sigma_{A_{-i}}(s_{-i}|s_i)$ is the conditional probability of the joint signal $(s_i, s_{-i})$ given the signal $s_i$, and $\sigma_{A_{-i}}(s_i)$ is the set of all conditional probabilities given $s_i$.*

Note that $\sigma_i$ is a probability distribution over $A_i$ while $\sigma_A$ is a probability distribution over $A$.

We assume that $\mathcal{M}$'s correlated strategy is public knowledge, but the actual instantiation, $s$, is not. In particular we assume that $\mathcal{M}$ sends each agent $i$ a private signal, $s_i$, based on $s$.

The agents are under no obligation to follow the mediator's signals. It is up to the mediator to pick a correlated strategy that a rational agent would be willing to follow. Note that our type of a mediator is different than Monderer and Tennenholtz's, where agents must agree to follow the mediator's suggested actions before knowing what they are [11].

**Definition 2.** *A correlated strategy $\sigma_A^* = \{\sigma_A(a)|a \in A\}$ is a correlated equilibrium if for every agent $i$ and every $s_i \in A_i$,*

$$\sum_{s_{-i} \in A_{-i}} \sigma_{A_{-i}}^*(s_{-i}|s_i) u_i(s_i, s_{-i}) \quad (2)$$
$$\geq \sum_{s_{-i} \in A_{-i}} \sigma_{A_{-i}}^*(s_{-i}|s_i) u_i(a_i', s_{-i}),$$

*for all $a_i' \in A_i$ [1]. The set of all correlated equilibria in $G$ is $C(G)$.*

If all of agent $i$'s opponents are following a correlated equilibrium $\sigma_A^*$, it is rational for agent $i$ to also follow $\sigma_A^*$.

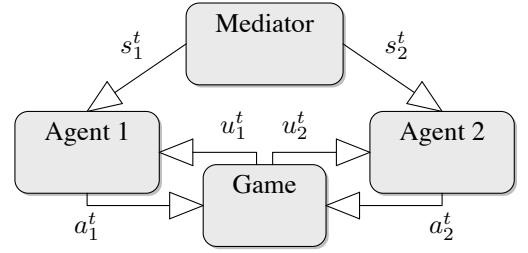

Figure 1: A graphical representation of our setting with 2 agents at time $t$.

In this paper, we are interested in a setting where agents have the ability to learn and adapt to the actions taken by others. Thus, we study repeated games. A repeated game $G^r = (G^1, G^2, \ldots)$ is an infinite sequence of the stage game $G$ played repeatedly. Agent $i$'s action at time $t$ is $a_i^t$ and the joint action at time $t$ is $a^t$. The history of joint actions, $hist(t) = \{a^1, \ldots, a^{t-1}\}$, is a record of the joint action taken at each iteration until time $t$. The empirical, or observed, percentage of play of joint actions, $\sigma_A^{hist(t)}$, is the percentage of time each joint action has been played as of time $t$. Agents may learn from previous iterations of the game to try and improve their strategy. Specifically, we assume that agent $i$ has a learning algorithm $L_i : hist(t) \to \Sigma_i$, that helps agent $i$ select a strategy for time $t$.

Let $\sigma_A^t$ be the actual correlated strategy at time $t$, *i.e.* the one agents are actually using and not necessarily the one based on $\mathcal{M}$'s suggestions. We say that $\sigma_A^t$ converges to a correlated equilibrium if for some $\sigma_A^* \in C(G)$, $\lim_{t \to \infty} \sigma_A^t = \sigma_A^*$. Thus, our algorithm is differentiated from algorithms that achieve convergence to the set of correlated equilibrium, for example [4, 8].

## 3 Setup

The setting for our paper is a repeated game $G^r$ with a mediator, $\mathcal{M}$. As illustrated for the two agent case in Figure 1, time $t$ will begin with the mediator giving each agent a suggested action, $s_i^t$. Agents will then simultaneously choose their action, $a_i^t$, which may or may not be $s_i^t$. If agent $i$ chooses not to follow $\mathcal{M}$'s signal, it can instead use a learning algorithm, $L_i$, which we assume is independent of $\mathcal{M}$'s signals, to select an action. Based on the actual joint action, each agent will then receive some utility and the process repeats. The mediator's signal to each agent is private information, known only to that agent and the mediator, as is the agent's utility function. However, the action set for each agent is public knowledge, as is the action taken by each agent during a turn.

The mediator's signals are based on a selected correlated strategy, $\sigma_A^{\mathcal{M}}$, which is constant throughout the repeated game. Although ideally the mediator will suggest a correlated strategy that is also a correlated equilibrium, each agent still needs to verify that the mediator has actually done so.

Our aim is to design an algorithm that achieves the following goals.

**First goal:** If $\sigma_A^{\mathcal{M}}$ is a correlated equilibrium then $\sigma_A^t$, the actual correlated strategy which is not necessarily $\sigma_A^{\mathcal{M}}$, will converge to $\sigma_A^{\mathcal{M}}$.

**Second goal:** If $\sigma_A^{\mathcal{M}}$ is not a correlated equilibrium, agents should be no worse off, in the limit, for having used our algorithm.

In Section 4, we present an algorithm, $\Lambda$, that achieves these goals with high probability. In Section 5, we generalize $\Lambda$ so that, with probability one, in the limit, it will achieve both goals. Since each agent will be using $\Lambda$ independently, we refer to $\Lambda_i$ as the instance of the algorithm being run by agent $i$ and $\Lambda$ as the joint algorithm.

The algorithm is based on the concept of giving $\mathcal{M}$ the benefit of the doubt; until there is reason to believe otherwise, agents assume that $\sigma_A^{\mathcal{M}}$ is a correlated equilibrium and follow $\mathcal{M}$'s signals. Specifically, agents will assume that the following conditions hold.

**Condition 1:** The correlated strategy $\sigma_A^{\mathcal{M}}$ is a correlated equilibrium.

**Condition 2:** All other agents are following the signals based on $\sigma_A^{\mathcal{M}}$.

Agents test whether these conditions hold during an initial period of play called a *sampling test* which has a fixed length of $l_T$. If, at the beginning of the sampling test, agent $i$ decides that one of the conditions does not hold, it will not follow $\mathcal{M}$'s signals and instead will use an individual "fall-back" strategy, $\gamma_i$, chosen uniformly at random. At the end of the sampling test, all agents who still believe that both conditions hold will continue to follow $\mathcal{M}$'s signals. All other agents will start using their original learning algorithm. The algorithm $\Lambda_i$ is correct if and only if, at the end of the sampling test, it correctly determines whether both conditions hold. The joint algorithm, $\Lambda$, is correct if and only if $\Lambda_i$ is correct for all $i$.

## 4 The Initial Algorithm

In this section, we describe how our initial algorithm works. As a first step in $\Lambda_i$, agent $i$ will check to see if Equation 2 holds for all $s_i \in A_i$. If Equation 2 does not hold, agent $i$ will know that Condition 1 cannot be true. In this case, agent $i$ will use a "fall-back" strategy, $\gamma_i \in \Sigma_i$, picked uniformly at random, for the rest of the sampling test. If Equation 2 does hold, agent $i$ must check to see if Condition 2 is true and will continue to follow $\mathcal{M}$'s signals throughout the sampling test.

Since the utilities for each agent, as well as the signals they receive each turn, are private, there may be no way to prove or disprove Condition 2 with absolute certainty at any finite point during the game. The best $\Lambda_i$ can do is reach a probabilistic conclusion. Since joint actions are public knowledge, $\Lambda_i$ can compare the empirical percentages of play for the duration of the sampling test against the percentages predicted by $\sigma_A^{\mathcal{M}}$. If the difference between these two values is statistically significant, there is a high probability that at least one agent has stopped following the mediator's signals.

To test if there is a difference, agent $i$ assumes there is some fixed but unknown correlated strategy $\tilde{\sigma}_A$ that all agents were actually using for the sampling test, where $\tilde{\sigma}_A$ may or may not be $\sigma_A^{\mathcal{M}}$. We are now able to use hypothesis testing, where our null hypothesis is that $\sigma_A^{\mathcal{M}}$ is equal to $\tilde{\sigma}_A$, *i.e.*,

$$H_0 : \sigma_A^{\mathcal{M}} = \tilde{\sigma}_A, \tag{3}$$

and our alternative hypothesis is that $\sigma_A^{\mathcal{M}}$ is not equal to $\tilde{\sigma}_A$, *i.e.*,

$$H_1 : \sigma_A^{\mathcal{M}} \neq \tilde{\sigma}_A. \tag{4}$$

The test statistic used is Pearson's $\chi^2$ test,

$$\mathcal{T} = \sum_{a \in A'} \frac{(X(a) - E(a))^2}{E(a)}, \tag{5}$$

where $A'$ is any subset of $A$ such that $|A'| = |A| - 1$, $X(a) = l_T \sigma_A^{hist(l_T)}(a)$ is the actual frequency of play of $a \in A'$ during the sampling test, $E(a) = l_T \sigma_A^{\mathcal{M}}(a)$ is the expected frequency of play according to $\sigma_A^{\mathcal{M}}$, and where $l_T$ is the length of the sampling period [12]. Note that $\sigma_A^{hist(l_T)}$ is based on a sampling from $\tilde{\sigma}_A$ of size $l_T$. For now we assume that $\sigma_A^{\mathcal{M}}(a) > 0$ for all $a \in A$. We relax this assumption later. The Pearson's $\chi^2$ test has (in the limit) a probability distribution function of

$$\chi^2_{df} + \chi^2_{NCP,1}, \tag{6}$$

where the first distribution has $df = |A| - 2$ degrees of freedom, and the second distribution has 1 degree of freedom and a non-centrality parameter of $NCP$ [9].

If $H_0$ is true, $NCP = 0$. Assuming that $H_0$ is true, we choose a significance level for rejection of the null hypothesis of $\alpha < 1$ and a corresponding critical value of $c(\alpha)$, *i.e.*, we reject the null hypothesis when $\mathcal{T} \geq c(\alpha)$. In this case, the probability of incorrectly rejecting $H_0$ (known as a Type 1 error) is $p_1 = \alpha$. If $H_1$ is actually true, we err when $\mathcal{T} < c(\alpha)$ and we do not reject $H_0$ (a Type 2 error). When $H_1$ is true, $NCP > 0$. Since the non-centrality parameter determines how much the probability distribution in Equation 6 gets adjusted, determining $NCP$ helps determine the probability of a Type 2 error.

The equation for $NCP$ is $NCP = t * \delta$, where $\delta$, the sensitivity parameter, is a measure of the difference between $\sigma_A^{\mathcal{M}}$ and $\tilde{\sigma}_A$ given by

$$\delta(\sigma_A^{\mathcal{M}}, \tilde{\sigma}_A) = \sum_{a \in A} \frac{(\tilde{\sigma}_A(a) - \sigma_A^{\mathcal{M}}(a))^2}{\sigma_A^{\mathcal{M}}(a)}. \tag{7}$$

For a given value of $\delta$, say $\hat{\delta}$, if

$$\delta(\sigma_A^{\mathcal{M}}, \tilde{\sigma}_A) \geq \hat{\delta}, \quad (8)$$

then the probability of a Type 2 error is bounded by some value $\beta(\hat{\delta}) < 1$, whose value is normally found via numerical computation [9]. Since $\beta$ is also a function of $l_T$ and $\alpha$, we refer to it as $\beta(l_T, \alpha, \delta)$.

Since agents do not know whether their opponents are following the mediator's suggestions, agents do not know the exact value for $\tilde{\sigma}_A$, and therefore, it is impossible to choose an appropriate value for $\hat{\delta}$ so that Equation 8 is guaranteed to hold. Instead, agents can consider a different question: what is the worst case situation under which Equation 8 does not hold? To answer this question, consider the set of all agents for whom Equation 2 does not hold, $N_B \subseteq N$. Let $(\sigma_{A_{-N_B}}^{\mathcal{M}}, \gamma_{N_B})$ be the actual correlated strategy for the duration of the sampling test, *i.e.*, a combination of those agents who will follow $\mathcal{M}$'s signals and those who will rely on their fall-back strategy. Let $\Sigma_{N_B}$ be the set of all possible joint strategies for agents in $N_B$, and

$$\Sigma_{N_B}(\sigma_A^{\mathcal{M}}, \delta) = \{\gamma_{N_B} \in \Sigma_{N_B} | \delta(\sigma_A^{\mathcal{M}}, (\sigma_{A_{-N_B}}^{\mathcal{M}}, \gamma_{N_B})) < \delta\} \quad (9)$$

be the set of all possible joint strategies for agents in $N_B$ which would result in Equation 8 not holding. Let $\mu(\Sigma_{N_B})$ and $\mu(\Sigma_{N_B}(\sigma_A^{\mathcal{M}}, \delta))$ be the Lebesgue measures of $\Sigma_{N_B}$ and $\Sigma_{N_B}(\sigma_A^{\mathcal{M}}, \delta)$, respectively. Then, since $\gamma_i$ is chosen uniformly at random, the probability of $\sigma_{N_B}$ being in $\Sigma_{N_B}(\sigma_A^{\mathcal{M}}, \delta)$ is

$$\psi(\Sigma_{N_B}) = \frac{\mu(\Sigma_{N_B}(\sigma_A^{\mathcal{M}}, \delta))}{\mu(\Sigma_{N_B})}. \quad (10)$$

Since agents do not know $N_B$, they consider the worst case scenario,

$$\psi = \max_{N' \subseteq N} \psi(\Sigma_{N'}). \quad (11)$$

If we assume that whenever Equation 8 does not hold and $\tilde{\sigma}_A \neq \sigma_A^{\mathcal{M}}$, a Type 2 error is always made, then the probability of a Type 2 error is at most

$$p_2 \leq (1 - \psi) \cdot \beta(\hat{\delta}) + \psi. \quad (12)$$

That is, Equation 8 holds with at least a probability of $\psi$ and when it does, the probability of a Type 2 error is at most $\beta(\hat{\delta})$ and with a probability of at most $\psi$, Equation 8 does not hold.

If we do not assume that $\sigma_A^{\mathcal{M}}(a) > 0$ for all $a \in A$, then Equations 5 and 7 may contain division by zero. To deal with this, we ignore all $a \in A$ such that $\sigma^{\mathcal{M}}(a) = 0$. If $\zeta = \{a \in A | \sigma^{\mathcal{M}}(a) = 0\}$, then the summations in Equations 5 and 7 need to exclude all $a \in \zeta$, and $df$ in Equation 6 now equals $|A| - 2 - |\zeta|$. If the null hypothesis is correct then $\sigma_A^{\mathcal{M}}(a) = 0$ implies that $\sigma_A^{hist(l_T)}(a) = 0$ for all $a \in \zeta$. Alternatively, if there exists $a' \in A$ such that $\sigma_A^{hist(l_T)}(a') > 0$ while $\sigma_A^{\mathcal{M}}(a') = 0$, the alternative hypothesis must be correct. Hence, both of these cases do not present problems.

The only other case is if for all $a \in A$ such that $\sigma_A^{\mathcal{M}}(a) = 0$, $\sigma_A^{hist(l_T)}(a) = 0$ but, unknown to the agents, the alternative hypothesis is correct. In this case, a Type 2 error may occur. To find the probability of this case happening, we first determine the probability of $a^t \in \zeta$. Since any agent who rejects $\mathcal{M}$'s suggested strategy chooses its new strategy uniformly at random, the probability, $\mathcal{P}$, that $a^t \in \zeta$ for $t \leq l_T$ is

$$\mathcal{P} \geq \sum_{a \in \zeta} \min_{N' \subseteq N} \sigma_{A_{-N'}}(a_{-N'}) \frac{1}{|A_{N'}|}, \quad (13)$$

where $\min_{N' \subseteq N}$ is considered since agents do not know $N_B$. Therefore, the probability that $a^t \notin \zeta$ for all $t \leq l_T$ is at most $(1 - \mathcal{P})^{l_T}$ and the overall probability of a Type 2 error is at most

$$p_2 \leq (1 - \mathcal{P})^{l_T} \left[ (1 - \psi) \cdot \beta + \psi \right]. \quad (14)$$

To accommodate the worst case, we assume equality holds in Equation 14. Note that $p_1$ has not changed. For simplicity, we assume that $p_1 = p_2 = p$, and refer to $p$ as the overall probability of error.

It is possible to rearrange $\beta(l_T, \alpha, \delta)$ to express $l_T$ as a function of $\alpha$, $\beta$ and $\delta$, *i.e* $l_T(\alpha, \beta, \delta)$. As a result, $l_T$ is the sample size needed to perform the test with at most a probability of error (of either Type 1 or Type 2) of $p$.

If all agents are to use the same value for $l_T$, they must also have the same value for $\beta$. This in turn requires them to have the same value for $\psi$. To achieve this, in Equations 11 and 13, agent $i$ will consider all possible $N'$, including those containing agent $i$.

### 4.1 Examples

In this section we provide two examples to illustrate how our test would work.

**Example 1:** Consider the game in Figure 2.

|  |  | Agent 2 | |
|---|---|---|---|
|  |  | $a_{2,1}$ | $a_{2,2}$ |
| Agent 1 | $a_{1,1}$ | 0,1 | 2,5 |
|  | $a_{1,2}$ | 5,2 | 1,0 |

Figure 2: A simple game

Let $A = \{(a_{1,1}, a_{2,1}), (a_{1,1}, a_{2,2}), (a_{2,1}, a_{2,1}), (a_{1,2}, a_{2,2})\}$. Suppose that $\mathcal{M}$ announces a correlated strategy, $\sigma_A^{\mathcal{M}} = \{1/18, 5/18, 2/18, 10/18\}$. Note that $\sigma_A^{\mathcal{M}}$ is a correlated equilibrium.

Suppose the agents choose $p = 0.1$ and $\delta = 0.01$. Agents must now determine the critical value for rejection, $c(\alpha)$,

and the length of the sampling test, $l_T$. Since $p_1 = \alpha$, $\alpha = 0.1$. For 3 degrees of freedom, $c(\alpha) = 6.25$. Since $\sigma_A^{\mathcal{M}}(a) > 0$ for all $a$, we can calculate $\beta$ by Equation 12. We calculate Equation 11 by numerical computation to find $\psi \approx 0.09429$. Therefore, $\beta = 0.0063$. In practice, $l_T(\alpha, \beta, \delta)$ would now be solved by some method of numerical computation [9]. For simplicity, we used the tables in Cohen to obtain a value of $l_T = 2100$ [2].

Suppose that after 2100 iterations, we have obtained an empirical frequency of play $\theta_A^{hist(2101)} = \{96, 601, 224, 1179\}$. Using Equation 5, we obtain a test statistic value of 4.678. Since this is lower than the critical value, both agents do not reject the null hypothesis and continue to use $\mathcal{M}$'s signals.

**Example 2:** Consider a different example based on the same game where $\mathcal{M}$ announces a correlated strategy of $\sigma_A^{\mathcal{M}} = \{2/18, 10/18, 1/18, 5/18\}$. In this case, $\sigma_A^{\mathcal{M}}$ is not a correlated equilibrium. Specifically, while Equation 2 is satisfied for Agent 1, it is not satisfied for Agent 2. Hence, Agent 2 will use a random fall-back strategy. Suppose $\gamma_2 = (3/4, 1/4)$.

For this example, the length of the test has not changed. Suppose we find an empirical frequency of $\theta_A^{hist(2101)} = \{1050, 350, 525, 175\}$ after 2100 turns. Since Agent 2 already knows that $\sigma_A^{\mathcal{M}}$ is not a correlated equilibrium, it will not perform the test. Agent 1 will obtain a test statistic value of 5953.3. This is well above the critical value and so Agent 1 will reject the null hypothesis, *i.e.*, it will stop following the signals of the mediator.

Note that, as we have stated our algorithm, Agent 1 will only know that there is a probability of at most 0.1 of incorrectly rejecting the null hypothesis. We have not accounted for the fact that the test statistic value is much higher than the critical value. An additional test that could be run after the null hypothesis is rejected is the calculation of the *p-value*. The p-value is the smallest $\alpha$ value that would still allow us to reject the hypothesis [12]. In the case of the above example, the p-value would be very small, and Agent 1 could be very certain that $\sigma_A^{\mathcal{M}}$ is not a correlated equilibrium.

## 5 Repeated Testing

The limitation of our basic test is that there is always some positive probability of error. This is due to the need to pick values for $1 - p$ and $\delta$ that are both greater than 0. Since we can pick any such values for $1 - p$ and $\delta$, this is not much of a practical limitation, however we may wish to achieve a stronger theoretical result. Our goal is to have agents converge to playing $\sigma_A^{\mathcal{M}}$ if it is a correlated equilibrium. If $\sigma_A^{\mathcal{M}}$ is not a correlated equilibrium, then the agents' utility should be no worse off for having used our algorithm. This leads to the idea of *repeated testing*, where throughout the repeated game, agents will use multiple iterations of $\Lambda_i$.

The set of repeated sampling tests is $R = \{R_1, R_2, \ldots\}$,

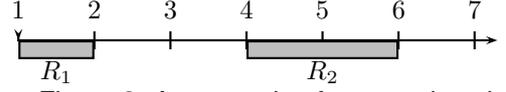

Figure 3: An example of repeated testing.

where $R_j = \{b_{R_j}, l_{R_j}\}$, $b_{R_j}$ is the first time period in $R_j$, and $l_{R_j}$ is the length of $R_j$. The instance of $\Lambda_i$ during test $R_j$ is denoted by $\Lambda_i^{R_j}$. The repeated tests are not contiguous. A simple example is shown in Figure 3, where the timeline represents a repeated game up to 7 iterations. The grey areas represent sampling test iterations. For example, $R_2 = \{b_{R_2}, l_{R_2}\} = \{4, 2\}$, meaning that the second test iteration begins at time period 4 and lasts for 2 iterations of the repeated game.

The parameters, $\delta$ and $p$, can be set to depend on the test iteration, *i.e.* $\delta(R_j)$ and $p(R_j)$. Each test period must be identical for each agent, *i.e.* $R_j$ must be the same for all agents. This means that $\delta(R_j)$ and $p(R_j)$ must be the same for all agents. The parameters are chosen such that

$$\lim_{j \to \infty} \delta(R_j) = 0, \qquad (15)$$

$$\sum_{j=1}^{\infty} p(R_j) < \infty. \qquad (16)$$

For example, we can let $\delta(R_j) = 1/j$ and $p(R_j) = 1/2^j$. Finally, we assume that each agent's fall-back strategy is fixed. That is $\gamma_i^{R_j} = \gamma_i^{R_{j'}}$, for all $j, j'$.

Our first result is that an agent will not draw the wrong conclusion about the mediator too often.

**Theorem 1.** *In the limit, with probability one, there will only be a finite number of tests where $\Lambda^{R_j}$ is incorrect.*

*Proof.* Let $\sigma_A^{\mathcal{M}}$ be the correlated strategy suggested by $\mathcal{M}$. Consider the following two cases:

$\sigma_A^{\mathcal{M}}$ **is a correlated equilibrium:** For test $R_j$, the probability of $\Lambda_i^{R_j}$ making a Type 1 error, $p_1(R_j)$, is equal to $p(R_j)$. By the Borel-Cantelli lemma, with probability one, there will only be a finite number of times $\Lambda_i^{R_j}$ is incorrect, *i.e.* makes a Type 1 error. [1] This reasoning can be applied to all agents, and therefore with probability one there will only be a finite number of times $\Lambda^{R_j}$ is incorrect.

$\sigma_A^{\mathcal{M}}$ **is not a correlated equilibrium:** If $\sigma_A^{\mathcal{M}}$ is not a correlated equilibrium, then some subset of agents, $N' \subseteq N$, will use their fall-back strategies instead of following the mediator's signals. The resulting correlated strategy for every test iteration will be $(\sigma_{A_{-N'}}^{\mathcal{M}}, \gamma_{N'})$.

Since $\gamma_{N'}$ is fixed, by Equation 15, there exists a finite $j^*$

---

[1] **Borel-Cantelli Lemma:** Let $\{E^t\}_0^\infty$ be a sequence of independent events and $P(E^t)$ be the probability of the event $E^t$ occurring. If $\sum_{t=0}^{\infty} P(E^t) < \infty$, then with probability one, only a finite number of the events will occur.

such that for all $j \geq j^*$,

$$\delta(\sigma_A^{\mathcal{M}}, (\sigma_{A_{-N'}}^{\mathcal{M}}, \gamma_{N'})) \geq \delta(R_j). \quad (17)$$

Let $\psi(R_j)$ be the value of $\psi$, according to Equation 11, during the sampling test $R_j$. Starting at $R_{j^*}$, we know that, with probability one, Equation 8 holds and therefore, since $\psi(R_j)$ is the probability of Equation 8 not holding, $\psi(R_j) = 0$, for all $j \geq j^*$. Therefore, the probability of a Type 2 error starting at $R_{j^*}$ is

$$p_2 = \sum_{j=j^*}^{\infty} (1 - \mathcal{P})^{l_T} \beta. \quad (18)$$

Note that $\mathcal{P}$, $l_T$ and $\beta$ are all functions $R_j$, however we omit the notation $(R_j)$ for clarity. Since $\beta$ is less than 1,

$$p_2 \leq \sum_{j=j^*}^{\infty} (1 - \mathcal{P})^{l_T} \left[ (1 - \psi) \cdot \beta + \psi \right] \quad (19)$$

$$= \sum_{j=j^*}^{\infty} p(R_j), \quad (20)$$

where $\psi$, as calculated by Equation 11, is also a function of $R_j$. Therefore, by Equation 16 and the Borel-Cantelli lemma, with probability one, there will only be a finite number of times $\Lambda_i^{R_j}$ is incorrect, *i.e.* makes a Type 2 error. Again, this reasoning can be generalized to all agents and therefore, there will only be a finite number of times $\Lambda^{R_j}$ is incorrect. □

We now examine the behaviour of agents between sampling tests. The periods between test iterations are called *free periods*. The set of free periods is $F = \{F_1, \ldots\}$ where $F_j = \{b_{F_j}, l_{F_j}\}$. Thus $G^r = \{R_1, F_1, R_2, F_2, \ldots\}$. For example, in Figure 3, the first free period, $F_1$, would be $\{b_{F_1}, l_{F_1}\} = \{2, 2\}$. If $\Lambda_i^{R_j}$ did not reject the null hypothesis, agent $i$ continues to follow $\mathcal{M}$'s signals for all of $F_j$. If $\Lambda_i^{R_j}$ did reject the null hypothesis, agent $i$ relies on its learning algorithm $L_i$ for $F_j$. We assume that $L_i$ is *flexible* at the beginning of each free period [3].

**Definition 3.** *The learning algorithm $L_i$ is flexible if at the beginning of every free period $F_j$,*

$$L_i(hist(b_{F_j})) = L_i(hist(1)). \quad (21)$$

*Therefore, during each free period, $L_i$ does not base its actions on what has happened before time $b_{F_j}$.*

For example, $L_i$ may be a trigger strategy, but that trigger may not be based on anything that has happened in a previous sampling test or free period.

We require that

$$\lim_{j \to \infty} \frac{\sum_{j'=1}^{j} l_{R_j}}{\sum_{j'=1}^{j} l_{F_j}} = 0, \quad (22)$$

for example $l_{F_j} = l_{R_j}^2$. This means that, in the limit, the length of the sampling periods is negligible compared to the length of the free periods. We also require that

$$\lim_{j \to \infty} \frac{l_{R_j}}{j} = \infty. \quad (23)$$

This means that the length of the sampling tests grows at faster than a linear rate. The specific values for $l_{R_j}$ and $l_{F_j}$ would have to be agreed upon by all agents.

**Definition 4.** *Let $\theta_A^{exp(t_1,t_2)}$ be the expected frequency of play from time $t_1$ to $t_2$, i.e., the expected number of times each joint action $a \in A$ gets played between times $t_1$ and $t_2$ inclusive. If $t_1$ is not given, we assume $t_1 = 1$. Similarly, let $\theta_A^{exp(F_j,\ldots,F_{j'})}$ be the expected frequency of play during the free periods $F_j$ through $F_{j'}$, inclusive.*

*Since the frequency of play depends on the algorithms the agents are using, let $\theta_A^{exp(t)}(L)$ be the expected frequency of play from time 1 to $t$ assuming that agents use the joint learning algorithm $L$ for the whole period.*

For simplicity in all of the following proofs, we assume that $t$ always corresponds to the beginning of a sampling period. Let $j(t)$ be the index of the last free period before $t$.

The first step is to show that if $\mathcal{M}$ suggests a correlated equilibrium, agents will converge to it.

**Theorem 2.** *If the correlated strategy suggested by $\mathcal{M}$, $\sigma_A^{\mathcal{M}}$, is a correlated equilibrium, then with probability one,*

$$\lim_{t \to \infty} \sigma_A^t = \sigma_A^{\mathcal{M}}. \quad (24)$$

*Proof.* If $\sigma_A^{\mathcal{M}}$ is a correlated equilibrium then by Theorem 1, with probability one, after some finite point $\Lambda$ will always correctly determine that $\sigma_A^{\mathcal{M}}$ is a correlated equilibrium. As a result, with probability one, after some finite point, all agents will choose to follow the mediator's signals during the free periods. □

Our next result is a technical lemma which shows that in the limit, agents are not harmed by taking time out to do the sampling tests.

**Lemma 1.** *In the limit, there is no difference between the average utility from agents using $L$ for the whole repeated game and just for the free periods, i.e.,*

$$\lim_{t \to \infty} \left[ u_i \left( \frac{\theta_A^{exp(t)}(L)}{t} \right) - u_i \left( \frac{\theta_A^{exp(F_1,\ldots,F_{j(t)})}(L)}{t} \right) \right] = 0. \quad (25)$$

*Furthermore, this is true even when excluding the first $j^* - 1$ free periods, for some $j^* > 1$, i.e.,*

$$\lim_{t \to \infty} \left[ u_i \left( \frac{\theta_A^{exp(t)}(L)}{t} \right) - u_i \left( \frac{\theta_A^{exp(F_{j^*},\ldots,F_{j(t)})}(L)}{t} \right) \right] = 0. \quad (26)$$

The proof is given in the Appendix.

Finally, we need to show that if $\sigma_A^{\mathcal{M}}$ is not a correlated equilibrium, agents are no worse off, on average, for having used $\Lambda$.

**Theorem 3.** *If the correlated strategy suggested by $\mathcal{M}$, $\sigma_A^{\mathcal{M}}$, is not a correlated equilibrium, then with probability one,*

$$\lim_{t\to\infty}\left[u_i\left(\frac{\theta_A^{exp(t)}(\Lambda)}{t}\right) - u_i\left(\frac{\theta_A^{exp(t)}(L)}{t}\right)\right] \geq 0. \quad (27)$$

*Therefore, in the limit, agent $i$ will be no worse off for using $\Lambda$ instead of $L_i$.*

*Proof.* If $\sigma_A^{\mathcal{M}}$ is not a correlated equilibrium, by Theorem 1, with probability one, starting at some sampling test, say $R_{j^*}$, $\Lambda$ will always correctly determine that $\sigma_A^{\mathcal{M}}$ is not a correlated equilibrium.

Consider $\theta_A$ with respect to some arbitary $a \in A$, denoted by $\theta_a$. We start by breaking the game down into the sequence of sampling tests and free periods. That is, $\theta_a^{exp(t)}(\Lambda) = \theta_a^{exp(R_1, F_1, \ldots, F(t))}(\Lambda)$. For $t \geq t(j^*)$, the utility can be split up into the utility for the sampling tests and free periods before $R_{j^*}$ and for those starting at $R_{j^*}$ i.e.,

$$\lim_{t\to\infty}\left[u_i\left(\frac{\theta_a^{exp(R_1, F_1, \ldots, R_{j^*-1}, F_{j^*-1})}(\Lambda)}{t}\right) + u_i\left(\frac{\theta_a^{exp(R_{j^*}, F_{j^*}, \ldots, F(t))}(\Lambda)}{t}\right)\right]$$

Since $\theta_A^{exp(R_1, F_1, \ldots, R_{j^*-1}, F_{j^*-1})}(\Lambda)$ is constant, in the limit, the first term is 0, and so we are interested in

$$\lim_{t\to\infty} u_i\left(\frac{\theta_a^{(R_{j^*}, F_{j^*}, \ldots, F(t))}(\Lambda)}{t}\right)$$

The expected frequency can be split up into the expected frequency for the sampling periods and for the free periods. Since $\Lambda$ always determines that $\sigma_A^{\mathcal{M}}$ is not a correlated equilibrium, during all the free periods agents will always use $L$, and so we are interested in

$$\lim_{t\to\infty}\left[u_i\left(\frac{\theta_a^{(R_{j^*}, \ldots, R(t))}(\Lambda)}{t}\right) + u_i\left(\frac{\theta_a^{(F_{j^*}, \ldots, F(t))}(L)}{t}\right)\right]$$

Since we assumed that all utilities are nonnegative, we may discard the first term, and thus have

$$\lim_{t\to\infty} u_i\left(\frac{\theta_a^{(F_{j^*}, \ldots, F(t))}(L))}{t}\right)$$

Therefore, by Lemma 1, the theorem follows. □

Together, Theorems 2 and 3 show that, with probability one, if $\sigma_A^{\mathcal{M}}$ is a correlated equilibrium, agents will converge to it and if $\sigma_A^{\mathcal{M}}$ is not a correlated equilibrium, agents will be no worse off in the long run for using $\Lambda$.

## 6 Conclusion

The setting for this paper was a repeated game with a mediator. The mediator makes suggestions to the agents as to what actions to take. We presented a test that agents could use so that, with high probability, they could determine if the mediator's suggestion was a correlated equilibrium. We then generalized our algorithm to incorporate repeated testing so that in the limit, with probability one, the test will always correctly determine whether the mediator's suggested strategy is a correlated equilibrium. As a result, if the mediator suggests a correlated equilibrium, then agents will converge to it, and otherwise, be no worse off in the long run for having used our algorithm.

We envision several directions for future research. First, it might be possible to extend our algorithm to work in radically uncoupled environments, where agents are not aware of the existence of others. This would significantly decrease the knowledge requirements of our test. Second, we would like to extend our approach so that the mediator receives feedback from the agents themselves, which can be used to help select appropriate correlated strategies. We believe that the incentive issues in such an approach will be challenging. It may also be interesting to apply our approach to other solution concepts such as mediated equilibria [11].

In a more applied direction, it might be possible to generalize our approach so it can be used in a stochastic game setting. Thus, our approach could be combined with methods such as Q-learning [7]. Correlated equilibria have also been used in graphical games, which can be used to model many different settings [10]. Hence, applying our technique to graphical games may yield some interesting results. For example, *network games* use graphical games to help represent a variety of problems, from public good provision and trade to information collection [6]. These models can be hindered by a "fundamental theoretical problem: even the simplest games played on networks have multiple equilibrium*[sic]* which display a bewildering range of possible outcomes" [6]. Our model may help integrate correlated equilibria as a possible solution to this problem.

## 7 Acknowledgements

Our thanks to Gord Hines for his statistical advice.

## References

[1] R. Aumann. Subjectivity and correlation in randomized strategies. *Journal of Mathematical Economics*, 1:67–96, 1974.

[2] J. Cohen. *Statistical Power Analysis for the Behavioral Sciences*. 2nd edition, 1988.

[3] D. P. D. Farias and N. Megiddo. Combining expert advice in reactive environments. *Journal of the ACM*, 53(5):762–799, 2006.

[4] D. P. Foster and R. Vohra. Calibrated learning and correlated equilibrium. *Games and Economic Behavior*, 21:40–55, 1997.

[5] D. P. Foster and H. P. Young. Learning, hypothesis testing, and Nash equilibrium. *Games and Economic Behavior*, 45:73–96, 2003.

[6] A. Galeotti, S. Goyal, M. O. Jackson, F. Vega-Redondo, and L. Yariv. Network games. Unpublished, Jan 2006.

[7] A. Greenwald and K. Hall. Correlated Q-learning. In *Proceedings of ICML-2003*, pages 242–249, Washington, DC, USA, 2003.

[8] S. Hart and A. Mas-Colell. A simple adaptive procedure leading to correlated equilibrium. *Econometrica*, 68:1127–1150, 2000.

[9] N. Johnson, S. Kotz, and N. Balakrishnan. *Continuous Univariate Distributions*, volume 2. 1995.

[10] S. Kakade, M. Kearns, J. Langford, and L. Ortiz. Correlated equilibria in graphical games. In *EC '03: Proceedings of the 4th ACM Conference on Electronic Commerce*, pages 42–47, New York, NY, USA, 2003.

[11] D. Monderer and M. Tennenholtz. Strong mediated equilibrium. In *Proceedings of the 21st American Association of Artificial Intelligence Conference*, Boston, MA, USA, 2006.

[12] L. Wasserman. *All of Statistics*. Springer, 2004.

## A  Proof of Lemma 1

*Proof.* Consider $\theta$ with respect to $a \in A$, denoted by $\theta_a$. Since $j^*$ is fixed, $\theta_a^{F_1,\ldots,F_{j^*-1}}(L)$ is constant, and therefore,

$$\lim_{t \to \infty} \frac{\theta_a^{exp(F_1,\ldots,F_{j^*-1})}(L)}{t} = 0, \qquad (28)$$

and therefore, Equations 25 and 26 are equivalent.

Since the utility functions are linear transformations, proving the following is sufficient, although not necessary, to prove that Equation 25 holds,

$$\lim_{t \to \infty} \frac{\theta_a^{exp(t)}(L) - \theta_a^{exp(F_1,\ldots,F_{j(t)})}(L)}{t} = 0. \qquad (29)$$

Since $L$ is flexible, it will, in expectation, always behave the same way during each free period. Specifically,

$$\theta_a^{exp(b_{F_j}, b_{F_j}+l_{F_j})}(L) = \theta_a^{exp(b_{F_{j'}}, b_{F_{j'}}+l_{F_{j'}})}(L), \qquad (30)$$

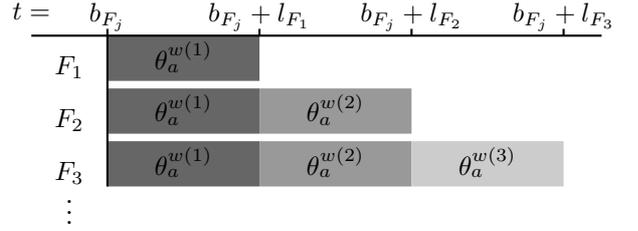

Figure 4: A graphical representation of how the expected frequency of play will be repeated each free period.

for all $j'$ such that $l_{F_{j'}} \geq l_{F_j}$. This relationship can be represented graphically, as shown in Figure 4, where for simplicity, we let $w(j) = exp(b_{F_j} + l_{F_{j-1}}, b_{F_j} + l_{F_j})$, where $l_{F_0} = 0$. Therefore,

$$\theta_a^{exp(F_1,\ldots,F_{j(t)})}(L) = \sum_{j=1}^{j(t)} (j(t) - j + 1)\theta_a^{w(j)}(L).$$

Note that $\theta_a^{w(j)}$ will be "represented" more than $\theta_a^{w(j')}$ for $j < j'$ and any finite $t$. In order for Equation 29 to hold, in the limit, all $\theta_a^{w(j)}$ be must represented equally, *i.e.*

$$\lim_{t \to \infty} \frac{j(t) - j + 1}{t} = \lim_{t \to \infty} \frac{j(t) - j' + 1}{t}, \qquad (31)$$

for all $j, j'$. Consider $t(j) = j^{-1}(t)$, *i.e.* the first time index after the $j^{th}$ free period has ended:

$$t(j) = \sum_{j'=1}^{j} (l_{R_j} + l_{F_j}) \geq \sum_{j'=1}^{j} l_{R_j}. \qquad (32)$$

By Equation 23, $\lim_{j \to \infty} \frac{t(j)}{j} = \infty$, and therefore,

$$\lim_{t \to \infty} \frac{j(t) - j + 1}{t} \leq \lim_{t \to \infty} \frac{j(t)}{t} = 0. \qquad (33)$$

Therefore, in the limit, all $\theta_a^{w(j')}$ will be represented equally. However, since $\sum_{j=1}^{j(t)} l_{F_j} < t$, each $\theta_a^{w(j)}$ will be "underrepresented" compared to $\theta_a^t(L)$ for any finite $t$. However, in the limit, this is not the case since,

$$\lim_{t \to \infty} \frac{\sum_{j=1}^{j(t)} l_{F_j}}{t} = \lim_{t \to \infty} \frac{\sum_{j=1}^{j(t)} l_{F_j}}{\sum_{j=1}^{j(t)} (l_{R_j} + l_{F_j})}$$

$$= \lim_{t \to \infty} \frac{1}{\frac{\sum_{j=1}^{j(t)} l_{R_j}}{\sum_{j=1}^{j(t)} l_{F_j}} + 1}$$

$$= 1 \text{ (by Equation 22)}. \qquad (34)$$

Therefore, in the limit $\theta_a^{w(j)}$ will be represented equally compared to $\theta_a^t(L)$. □